\documentclass[a4paper,12pt]{elsarticle}
\usepackage{amsmath}
\DeclareGraphicsExtensions{.eps}
\usepackage{indentfirst}
\usepackage{amsthm}
\usepackage{rotating}
\begin{document}
\begin{frontmatter}
\author{Demetris P.K. Ghikas\corref{cor1}}
\ead{ghikas@physics.upatras.gr, Tel: +302610-997460, FAX: +302610-997617}
\author{Athanasios C. Tzemos}
\address{Department of Physics University of Patras,\\ Patras 26500, Greece}
\cortext[cor1]{Corresponding author}
\title{Entangling and decohering influence of noisy perturbations. Stochastic anti-resonance in the time evolution of interacting qubits.}
\begin{abstract}
We study the entanglement evolution of two coupled qubits in interaction with an external environment and in the presence of an external magnetic field with a stochastic component. The results show the expected degradation of entanglement due to the noise. The new effect is that the time of disentanglement depends in a non-monotonous way on the strength of the noise. We find that it is shortest for an intermediate strength value of the latter. This we call "stochastic anti-resonance". Our results could lead to a better undestanding of noisy perturbations and their role for optimal designing of quantum devices.
\end{abstract}
\begin{keyword}{entanglement, Heisenberg chain, noise, stochastic resonance}
\end{keyword}
\end{frontmatter}
\section{Introduction}
\indent
The source of difficulties for the scalability of quantum information processing devices is the constant interaction with their local environments \cite{Casati,Nielsen,Preskill,Schlosshauer,Hornberger}. Engineering these interactions can improve the performance, but residual noisy agents, like non-zero temperature environments, or noisy external fields of control, impair the optimality of the behavior. Thus, appart from the well established methods of decoupling schemes \cite{Viola}, error correcting codes \cite{Knill} and decoherence free subspaces \cite{Lidar}, addressing directly the role of noise is an extra possibility for the engineering methods. The discovery of the counterintuitive phenomenon of stochastic resonance, in both classical non-linear \cite{Chapeau} and quantum systems \cite{Grammaitoni}, points to the direction of a non trivial  role of noise as a (possibly beneficial) control. There are extensive studies of entanglement and decoherence in thermal environments and in the presence of external magnetic fields and noise \cite{Wang, Zhou,Rigolin,Kamta,Shan,Yeo,Meng}. Especially the stochastic control and stochastic resonance has been studied in various arrangements   \cite{Altafini, Mancini,Wodkiewicz,Wild, Rivas01}. Here we study the dynamical evolution of entanglement for a two qubit Heisenberg chain in interaction with an external bosonic environment and a noisy external magnetic field. We are interested in the evolution of the effects of classical noise on the quantum properties of a simple system that can be solved analytically. More specifically, we have looked for a possible non-monotonous dependence of some physical quantity on the strength of an external noise, namely  whether we can have a stochastic resonance type behavior. The appearance of such a behavior, appart from its inherent interest, would be useful for a better understanding of noisy perturbations in real quantum information devices. What we have found is that the noise degrades initial entanglement, something expected, but the time taken for this effect is not a monotonous function of the noise strength for some initial states. This is interesting since it shows that an intermediate strength of noise can be worse than a smaller or bigger. Since we observe a worse degradation, we call the effect stochastic antiresonance. What in fact we have found is that the Entanglement Sudden Death (ESD) time \cite{eber1,eber2,eber3,eber4,eber5,eber6} does not depend monotonically on the strength of the noise. We are currently looking to multiqubit systems with a wider range of parameter values and with the use of stochastic differential equations for a direct numerical study. In the paper first we introduce the corresponding master equation where the external gaussian white classical noise is incorporated in terms of a double commutator. Then the Heisenberg type Hamiltonian is presented with some comments on the parameter values. Our results on the concurrence and purity evolution are presented in the form of 2d and 3d diagrams where the appearance of stochastic antiresonance effect is made evident. 
\section{Quantum systems driven by external white noise}
\indent
In order to be able to have analytic expressions for the dynamics, we use a form of the master equation where the noise is incorporated in terms of an appropriate term as in \cite{Luczka}. For a zero temperature bath the master equation is
\begin{align}
\frac{d\rho_s}{dt}=-i[H,\rho_s]+\gamma D[S^-_1]\rho_s+\gamma D[S^-_2]\rho_s-M[V,[V,\rho_s]]\label{eksisosi},
\end{align}
where V=$S_1^z+S_2^z$, $ D[S^-_1]\rho_s=S^-_1\rho_s(S^-_1)^{\dagger}-\{(S^-_1)^{\dagger}S^-_1,\rho_s\}/2$, $D[S^-_2]\rho_s=S^-_2\rho_s(S^-_2)^{\dagger}-\{(S^-_2)^{\dagger}S^-_2,\rho_s\}/2$ and $\gamma$ is the rate of population relaxation. We assumed that each qubit has the same interaction with the environment.
The first term describes the unitary evolution of the system, the second and third terms the interaction between the system and a thermal environment. The last term is the addition of an external classical gaussian white noise \cite{Luczka}.

\section{Two qubit XY Heisenberg model}
\indent
The general form of a N-spin Heisenberg chain (for spin $1/2$ particles) with nearest-neighbor interaction is

\begin{align}
H=\sum_{n=1}^N\Big(J_xS_n^xS_{n+1}^x+J_yS_n^yS_{n+1}^y+J_zS_n^zS_{n+1}^z\Big),
\end{align}
where $S_n^i=\frac{1}{2}\sigma^n_i(i=x,y,z)$ the spin-$1/2$ operators, $\sigma^n_i$ the corresponding Pauli's operators and $\hbar=1$.  Moreover if the chain is periodic, the boundary condition is $S_{N+1}=S_1$. The chain is called XYZ model for arbitrary values of $J_i$. If $J_i<0$ the chain is called ferromagnetic and if $J_i>0$, antiferromagnetic. Here we study the case where $J_z=0$. The Hamiltonian of a two qubit Heisenberg XY chain in the presence of an external magnetic field $\omega$ is \cite{Wang}:
\begin{align}
H=\omega(S^z_1+S_2^z)+J(S^+_1S^-_2+S^-_1S_2^+)+\Delta(S^+_1S_2^++S^-_1S^-_2)\label{Hamiltonian},
\end{align}
where $J=(J_x+J_y)/2$, $\Delta=(J_x-J_y)/2$ and $S^{\pm}=S^x\pm iS^y$. The first term describes the energy levels of spins in the external field, while the other two terms are the spin-interaction Hamiltonian and describe the coherence production of the two qubits.

As it has been shown \cite{Wang},\cite{Shan} the third term is essential for the production of a non-vanishing steady-state entanglement even in the presence of decoherence. Moreover, one can produce entanglement for any finite temperature, by adjusting the magnetic field, but only in the anisotropic case ($\Delta\neq 0$) \cite{Kamta}. The XY Heisenberg chain has been studied intensively in the last decade, for various purposes, such as NMR quantum computation \cite{Chuang} and quantum teleportation \cite{Yeo}, because of its simplicity (especially in the two-qubit case) and its many interesting features.

For the sake of simplicity we assumed that $\gamma$ is the same for both qubits. This assumption requires the use of system parameters, such that the inequalities \begin{align}\frac{\sqrt{\omega^2+\Delta^2}-\omega}{\omega}\leq0.1\end{align} and \begin{align}\frac{|J|}{\omega}\leq0.1\end{align} are satisfied. These equalities ensure that the interaction between the two qubits does not alter the energy level separations
by more than $10\%$ from that of the non-interacting system.\cite{Wang} All of our examples satisfy the above equalities.

\section{Concurrence and purity evolution for $T=0$}
\indent
We use concurrence  for the quantification of entanglement \cite{Horodecki},\cite{Kus},\cite{Wooters},\cite{Wooters2}. For a system  density matrix $\rho$, the concurrence $C$ is

\begin{align}
C=\max(\sqrt{\lambda_1}-\sqrt{\lambda_2}-\sqrt{\lambda_3}-\sqrt{\lambda_4},0),
\end{align}
where $\lambda_1,\lambda_2,\lambda_3,\lambda_4$ are the eigenvalues of spin flipped density matrix R (with $\lambda_1$ the largest one). The definition is:
\begin{align}
R=\rho(\sigma_y\otimes \sigma_y)\rho^*(\sigma_y\otimes \sigma_y)
\end{align}
$C$ is in the range $[0,1]$. $C=0$ corresponds to a product state, while $C=1$ to a maximally entangled state. All the other states inside this range are called partially entangled.

For the purity of quantum states, we use the linear entropy function, defined as:
\begin{align}
L=1-Tr(\rho^2)
\end{align}

The basis states for our system are the usual product states:
\begin{align}
&|\psi_1\rangle=|e\rangle_1\otimes |e\rangle_2,\\&
|\psi_2\rangle=|e\rangle_1\otimes |g\rangle_2,\\&
|\psi_3\rangle=|g\rangle_1\otimes |e\rangle_2,\\&
|\psi_4\rangle=|g\rangle_1\otimes |g\rangle_2.
\end{align}

A simplifying feature of  Hamiltonian \eqref{Hamiltonian} is that the dynamics of  the system evolution splits into two independent sets of equations for the  two submatrices.
\begin{align}
\begin{pmatrix}
\bullet&\bullet&\bullet&\bullet\\ \bullet&\bullet&\bullet&\bullet\\ \bullet&\bullet&\bullet&\bullet\\ \bullet&\bullet&\bullet&\bullet
\end{pmatrix}\to \begin{pmatrix}
\bullet&0&0&\bullet\\ 0&\bullet&\bullet&0\\ 0&\bullet&\bullet&0\\ \bullet&0&0&\bullet
\end{pmatrix}+\begin{pmatrix}
0&\bullet&\bullet&0\\ \bullet&0&0&\bullet\\ \bullet&0&0&\bullet\\ 0&\bullet&\bullet&0\label{k1}
\end{pmatrix}
\end{align}
We note that the second submatrix has not the form of a density matrix. In this paper we use the first submatrix for our results, which can describe product states, entangled states and mixtures of them. For such a density matrix, the concurrence and linear entropy functions are found to be equal to:
\begin{align}
C=\max\{0,C_1,C_2\},
\end{align}
where
\begin{align}
&C_1=2(|\rho_{41}|-\sqrt{\rho_{33}\rho_{22}}),\\&
C_2=2(|\rho_{32}|-\sqrt{\rho_{44}\rho_{11}}),
\end{align}
and
\begin{align}
L=1-\sum_{i=1}^4\rho_{ii}^2(t)-2|\rho_{32}(t)|^2-2|\rho_{41}(t)|^2
\end{align}
correspondingly.

We present the steady state solution analytically in the appendix. Here we report on  our results with 2-d and 3-d diagrams corresponding to some fixed values of system parameters.  For $M=0$ we get the results of \cite{Wang}. In all cases the parameter values are $J=\Delta=0.1, \gamma=0.01$ and $\omega=1$.\footnote{$\gamma\ll\omega$ so that the weak coupling approximation remains valid.}
We have studied the evolution for various  values of the noise strength, with initial conditions associated with product states, Bell states and mixed states. The corresponding representation of the initial density submatrices are as follows.
\\
\textit{Product state} $|gg\rangle$:
\begin{align}
\begin{pmatrix}
0&0&0&0\\ 0&0&0&0\\ 0&0&0&0\\ 0&0&0&1
\end{pmatrix}
\end{align}
\\
\textit{Product state} $|ee\rangle$:
\begin{align}
\begin{pmatrix}
1&0&0&0\\ 0&0&0&0\\ 0&0&0&0\\ 0&0&0&0
\end{pmatrix}
\end{align}
\\
\textit{Bell state} $\frac{|ee\rangle+|gg\rangle}{\sqrt2}$:
\begin{align}
\begin{pmatrix}
1/2&0&0&1/2\\ 0&0&0&0\\ 0&0&0&0\\ 1/2&0&0&1/2
\end{pmatrix}
\end{align}
All of the above initial conditions have matrix representation of the form
\begin{align}
\begin{pmatrix}
\bullet&0&0&\bullet\\ 0&0&0&0\\ 0&0&0&0\\ \bullet&0&0&\bullet
\end{pmatrix}
\end{align}
(i.e. the evolution begins outside the central subspace $\{|eg\rangle,|ge\rangle\}$)
and do not exhibit stochastic antiresonance. Noise degrades entanglement, both the initial and final, in a monotonous way.
\begin{figure}
    \centering
\scalebox{0.33}{\includegraphics[angle=0]{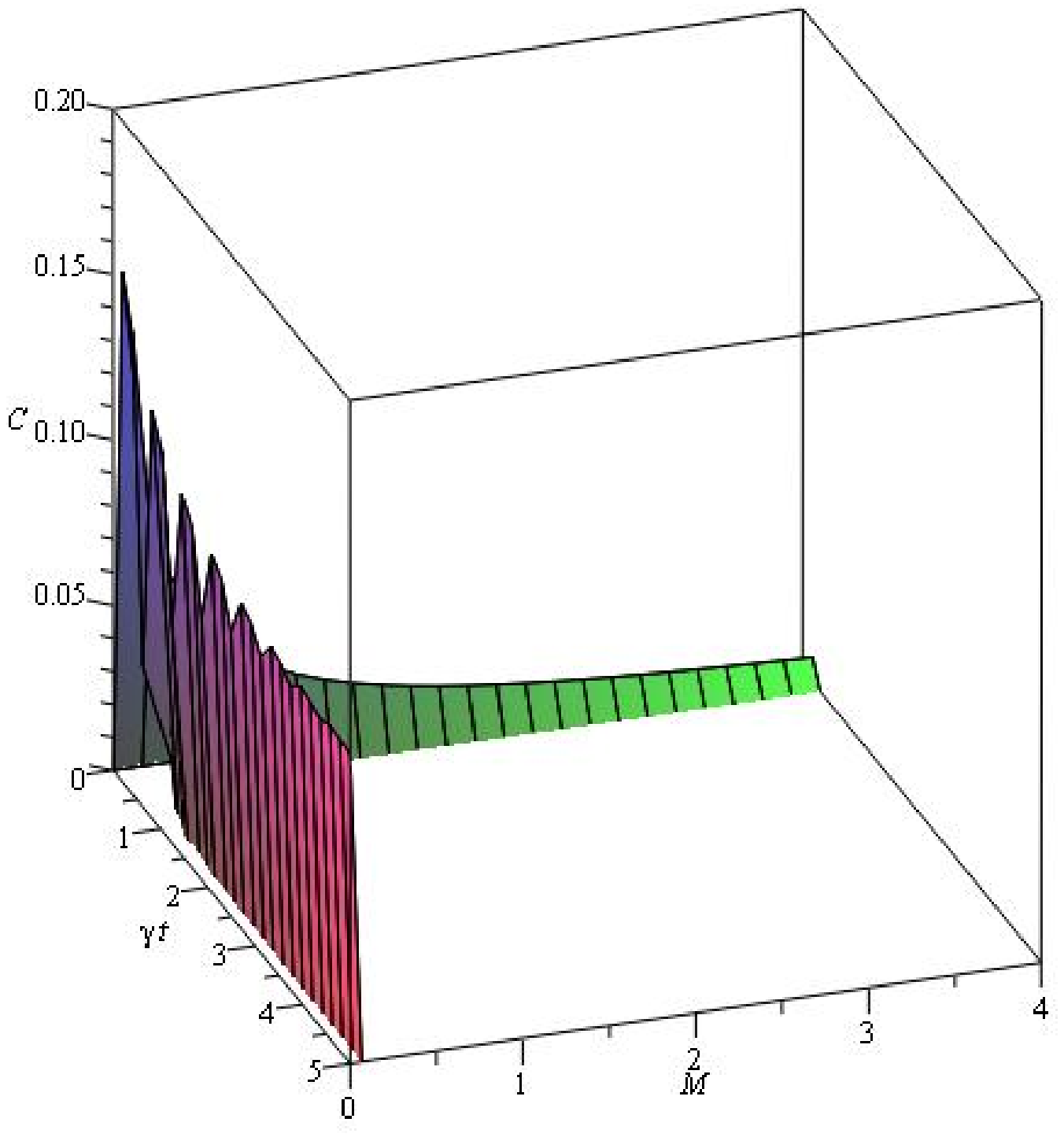}}
\scalebox{0.33}{\includegraphics[angle=0]{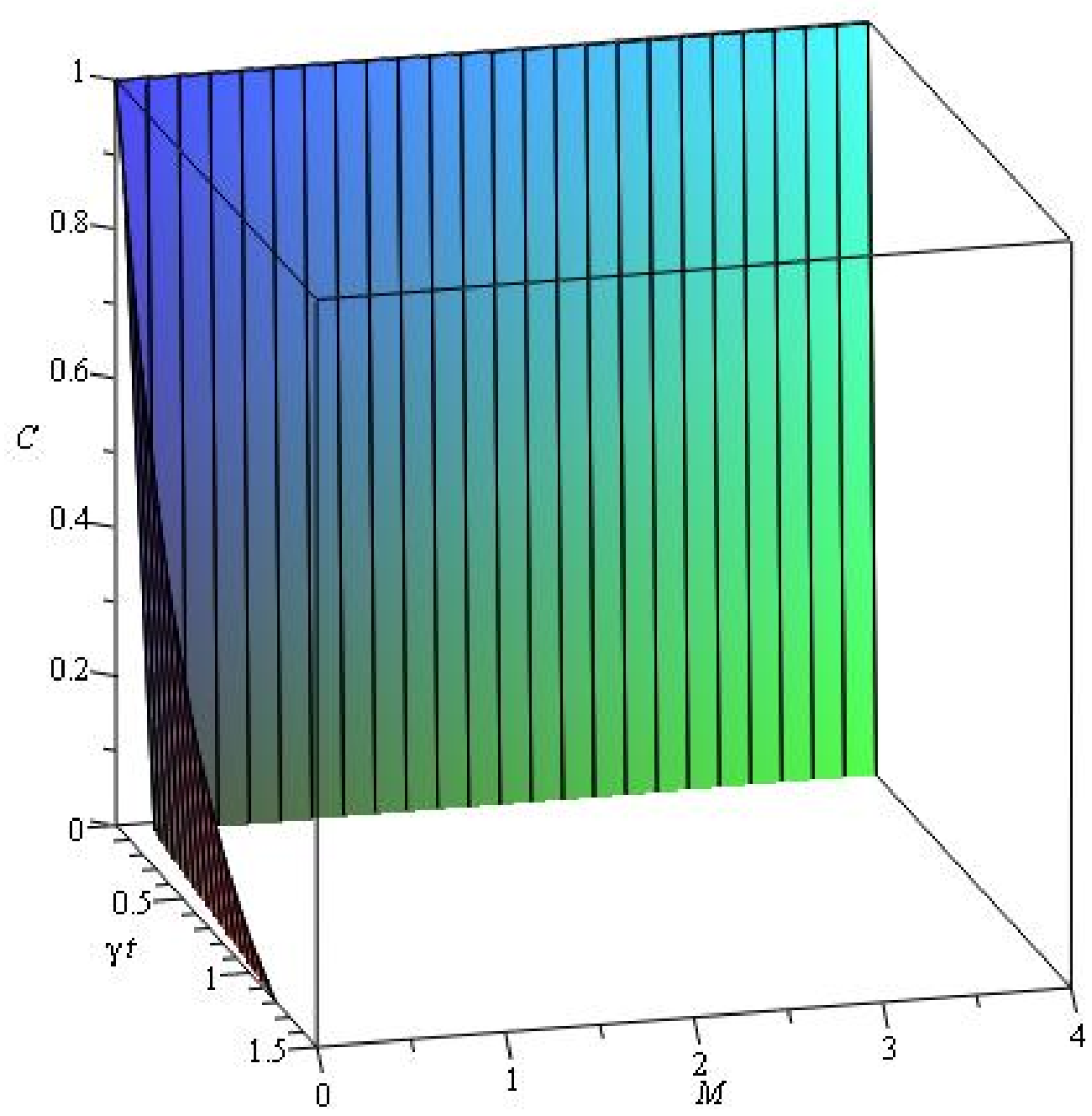}}
\caption{Concurrence as a function of time and noise parameter M for the product state $\psi_{in}=|gg\rangle$ (left) and Bell state $\psi_{in}=\frac{|ee\rangle+|gg\rangle}{\sqrt2}$ (right) ($J=\Delta=0.1, \gamma=0.01$ and $\omega=1$). Noise degrades entanglement monotonously.} \label{gg and bell state ee+gg concurrence} \end{figure}

\begin{figure}[htp]
	\centering
\scalebox{0.33}{\includegraphics[angle=0]{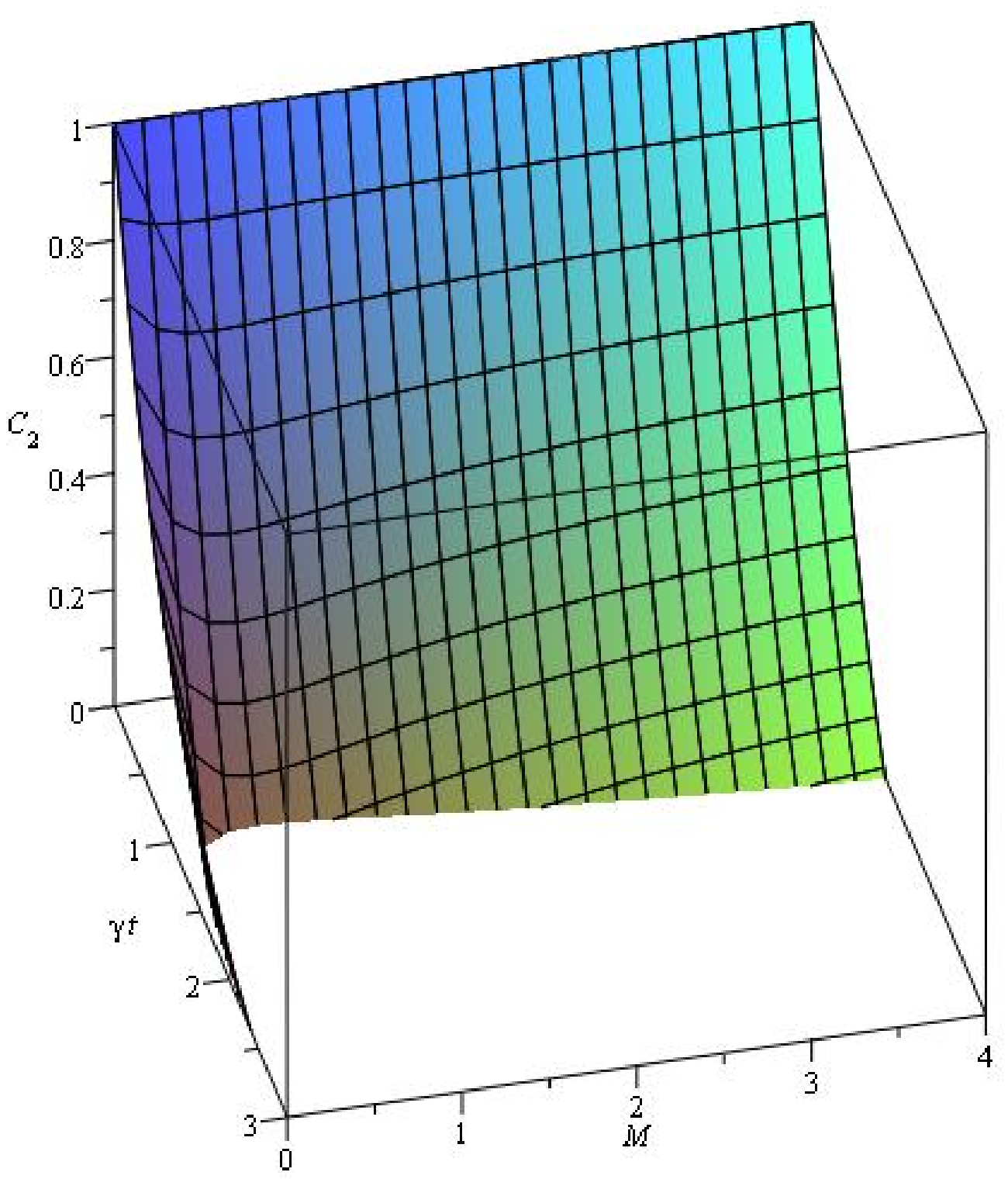}}
\scalebox{0.33}{\includegraphics[angle=0]{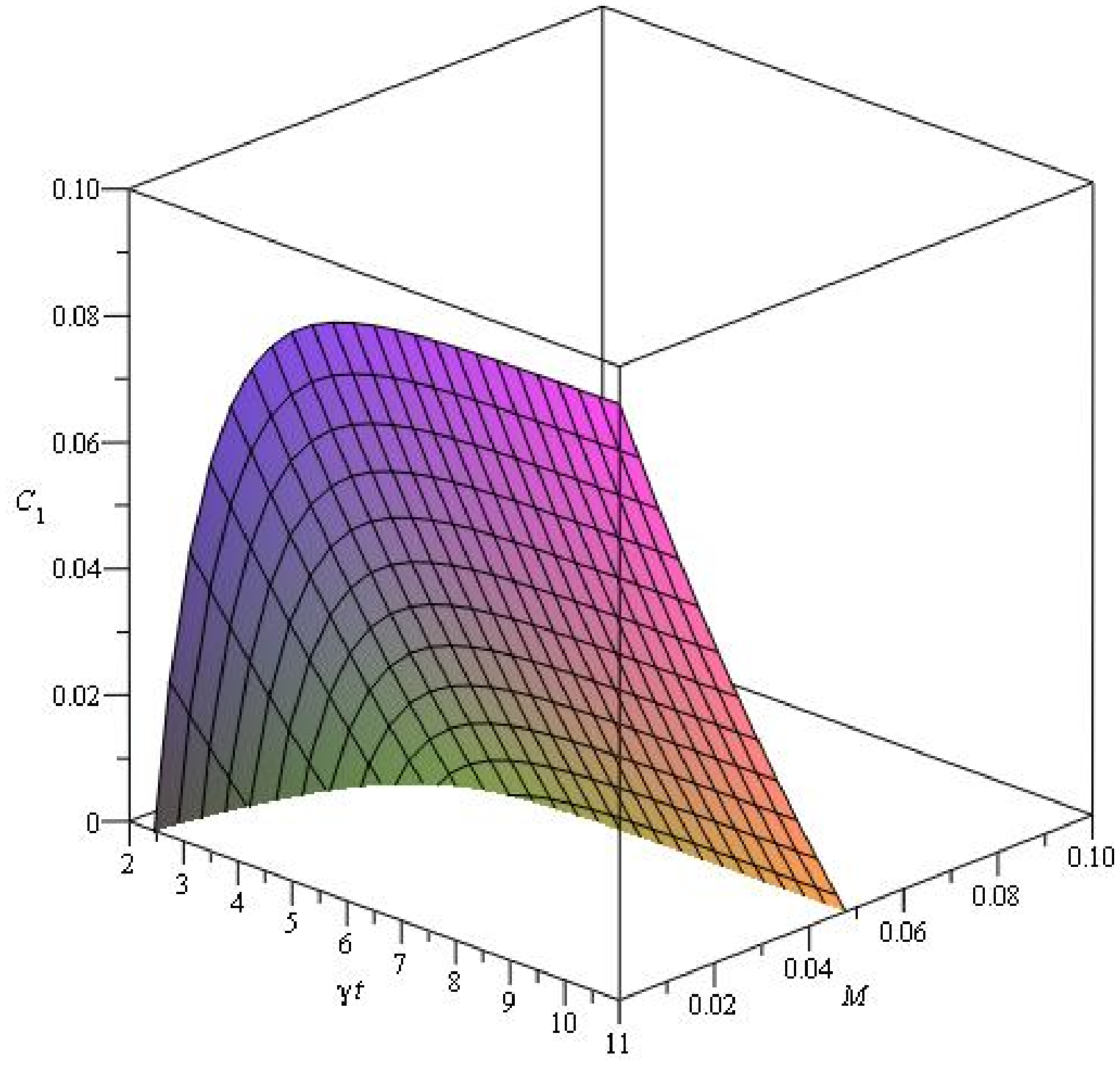}}
	\caption{Concurrence collapse (left) and revival (right) as a function of time and noise parameter M for the Bell state $\psi_{in}=\frac{|eg\rangle+|ge\rangle}{\sqrt2}$. On the left it is clear that there is a noise parameter value near $0.5$ for which the collapse time is fastest. On the right it is obvious that the bigger the noise parameter the smaller the steady state entanglement becomes. Parameters as in Figure 1. }
	\label{fig:Bell_stae_collapse_revival}
\end{figure}
\begin{figure}[t] \centering
\scalebox{0.70}{\includegraphics[angle=0,
width=1\textwidth]{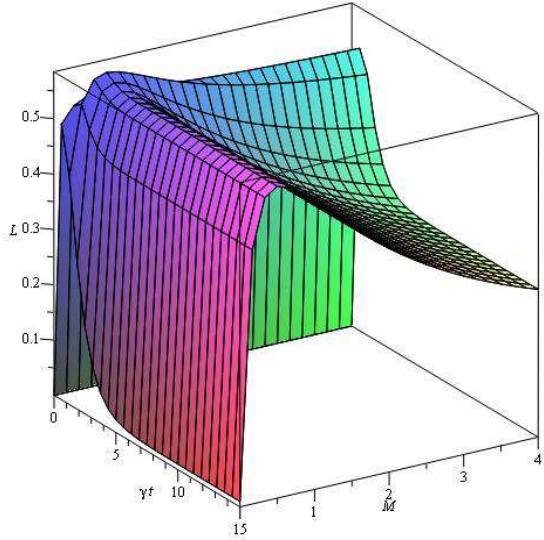}}
\caption{Linear entropy as a function of time and noise parameter M for Bell state $\psi_{in}=\frac{|eg\rangle+|ge\rangle }{\sqrt2}$. Parameters as in Figure 1. } \label{Bell_state_linear _entropy} \end{figure}
\begin{figure}
	\centering
\scalebox{0.25}{\includegraphics[angle=0]{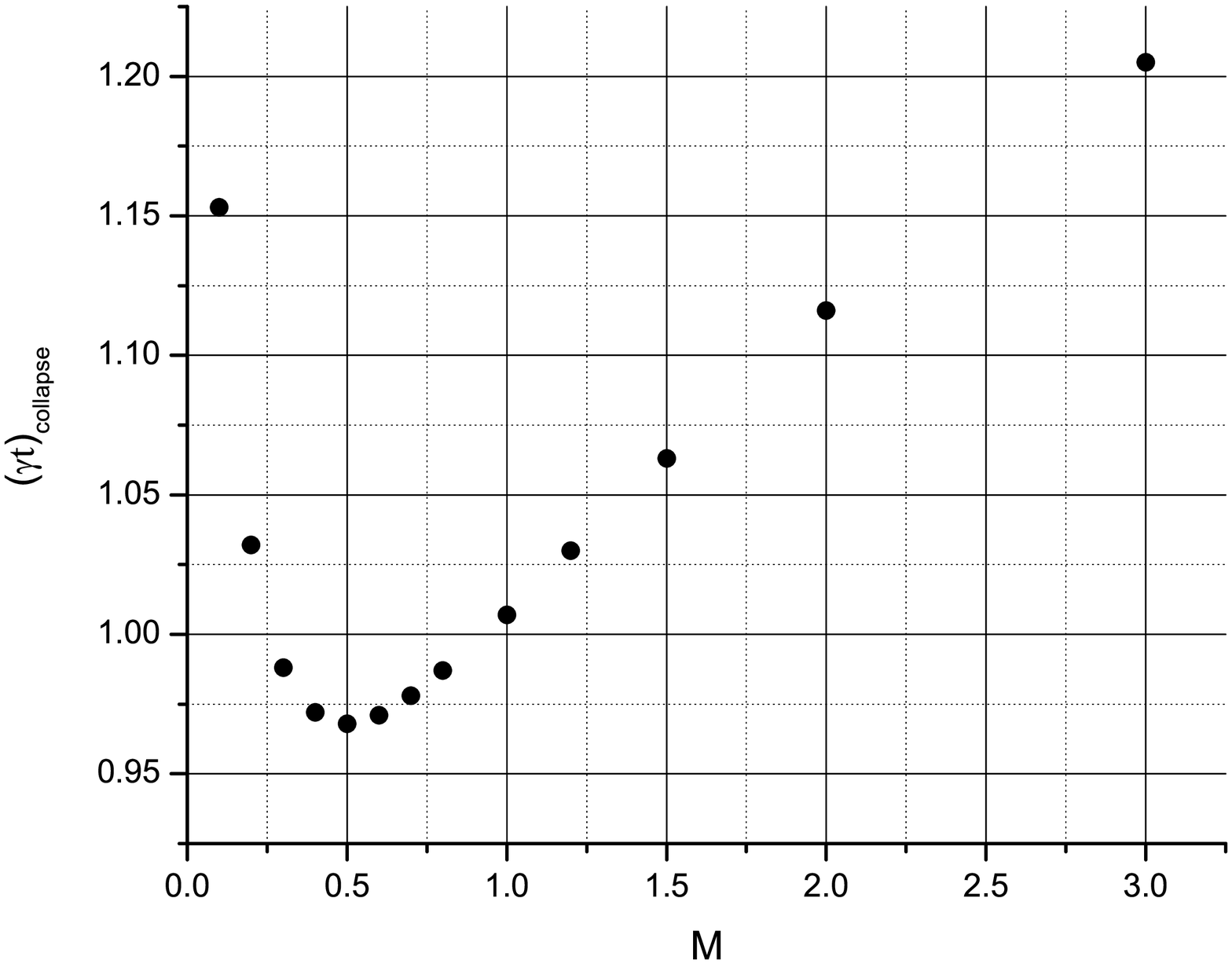}}
\scalebox{0.25}{\includegraphics[angle=0]{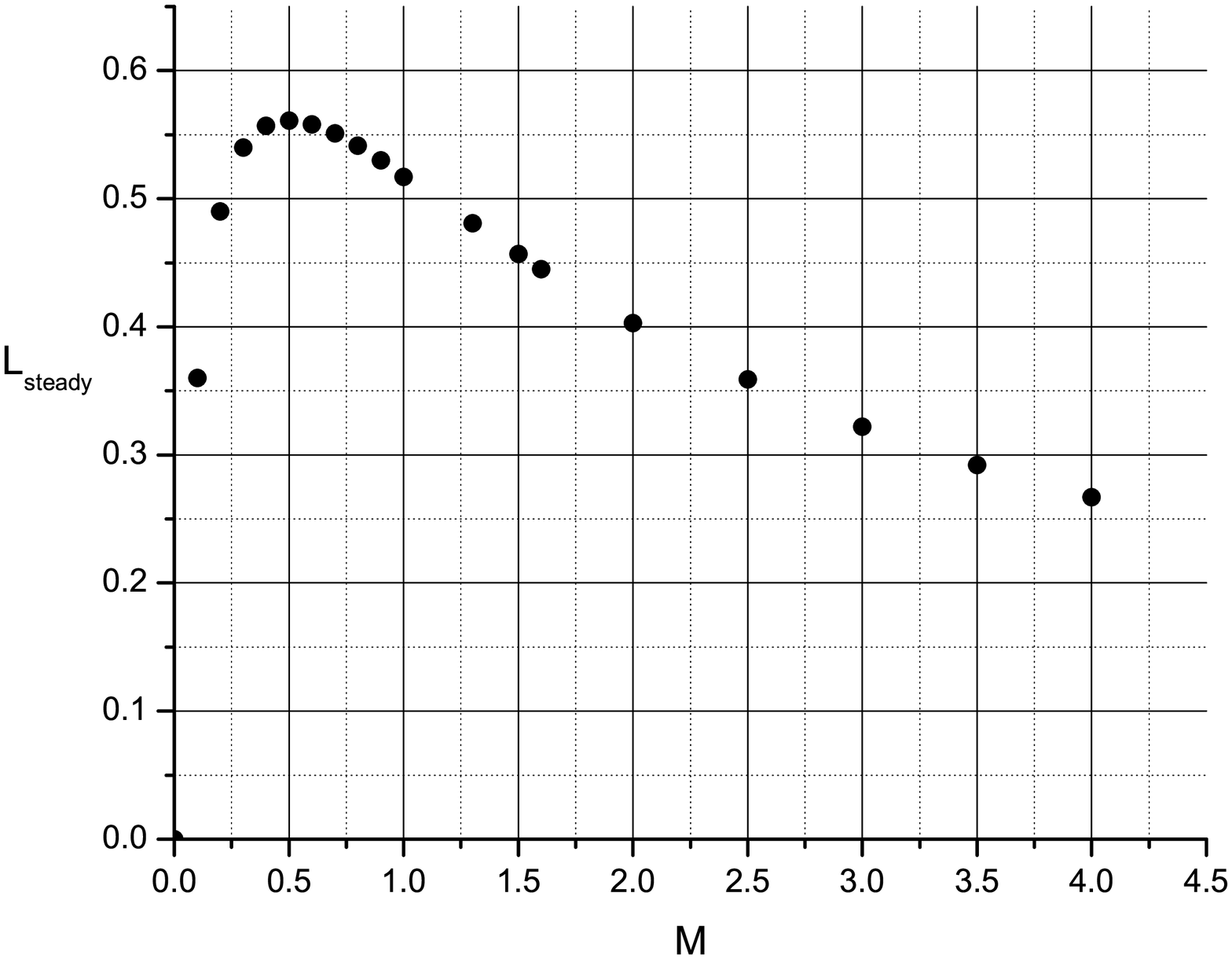}}
\scalebox{0.25}{\includegraphics[angle=0]{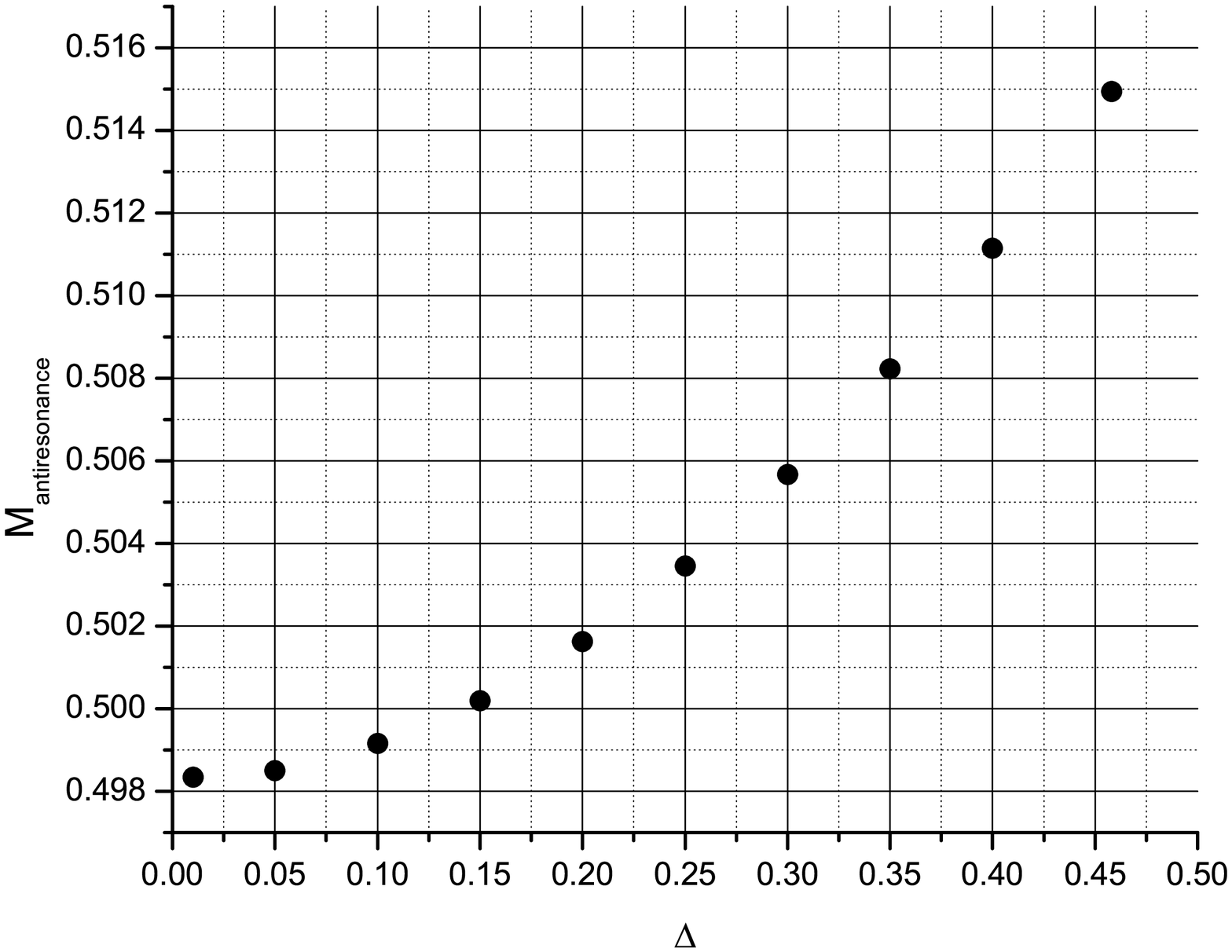}}
\scalebox{0.25}{\includegraphics[angle=0]{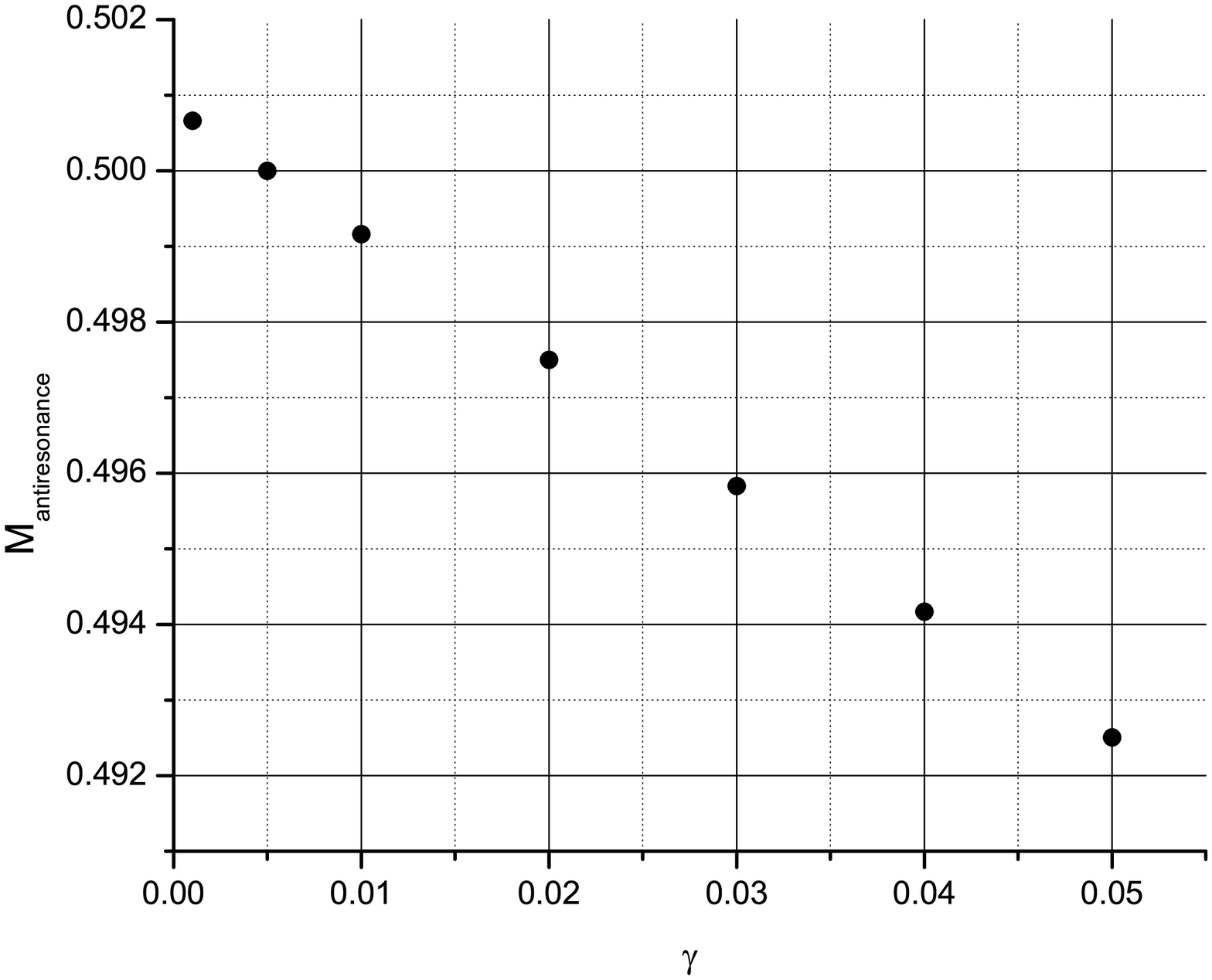}}
	\caption{Sudden death time (upper left) and steady mixedness (upper right) as a function of noise parameter M for the Bell state $\psi_{in}=\frac{|eg\rangle+|ge\rangle }{\sqrt2}$. Parameters as in Figure 1. The non monotonous dependence on M is obvious. $M_{antiresonance}$ as a function of anisotropy $\Delta$ (lower left) and population relaxation rate $\gamma$ (lower right). We see that both $\Delta$ and $\gamma$ have a small influence on $M_{antiresonance}$. }
	\label{fig:non monotonicity_01}
\end{figure}
\begin{figure}
	\centering
\scalebox{0.33}{\includegraphics[angle=0]{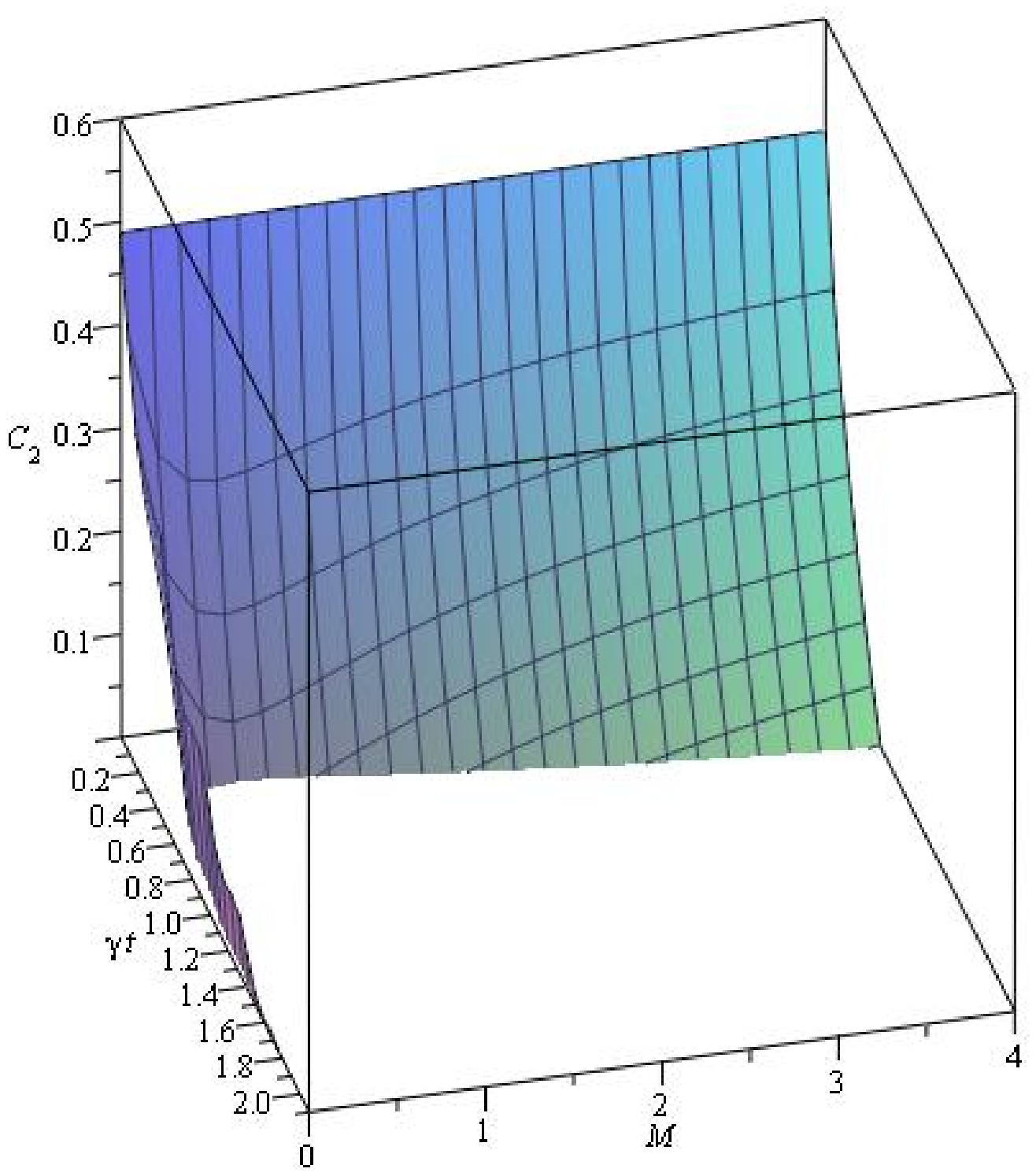}}
\scalebox{0.33}{\includegraphics[angle=0]{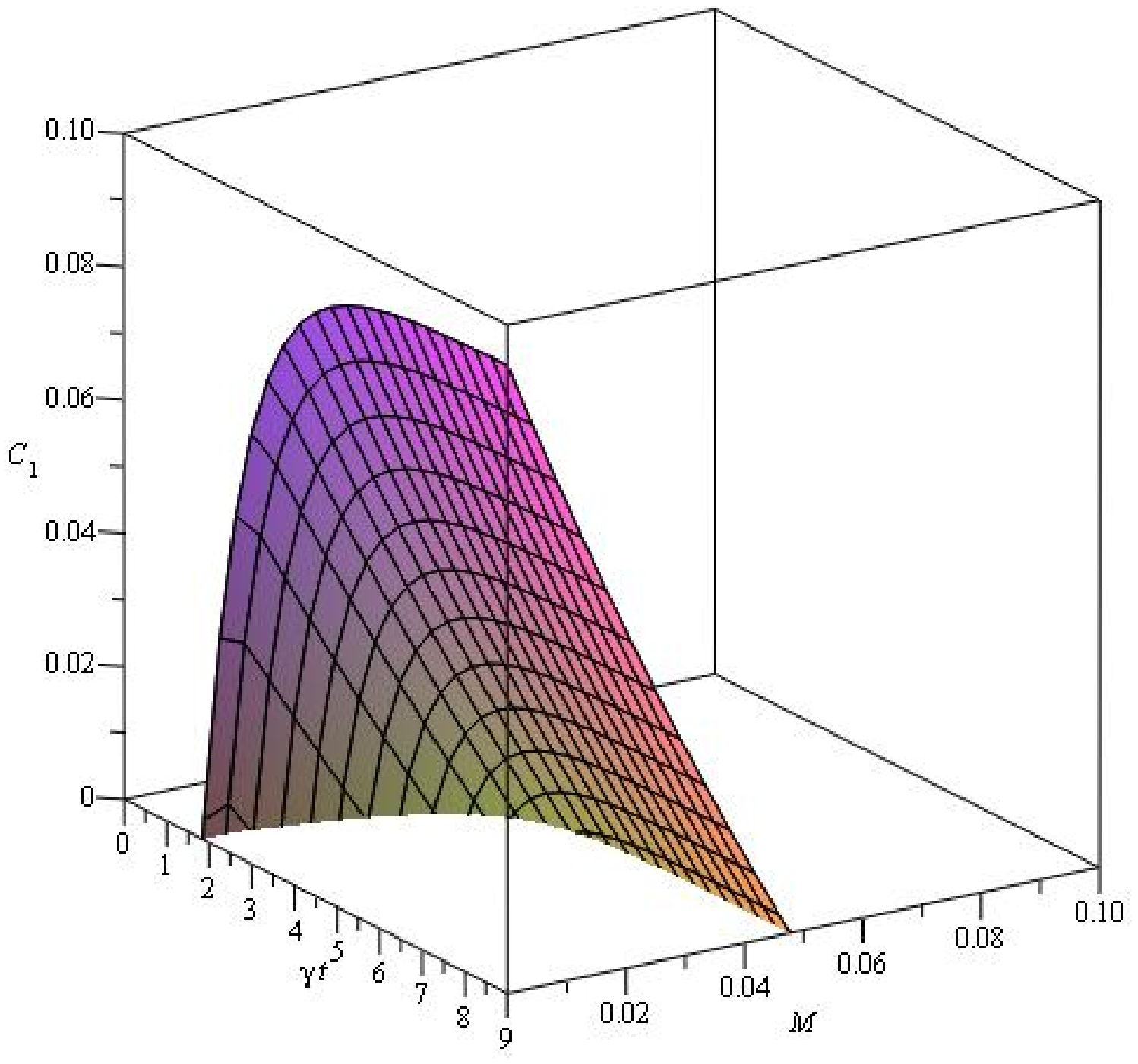}}
	\caption{Concurrence collapse (left) and revival (right)as a function of time and noise parameter M for mixed state $\rho_{in}=\frac{1}{2}|gg\rangle\langle gg|+\frac{1}{2}(\frac{(|{eg}\rangle+|ge\rangle)(\langle{eg}|+\langle ge|)}{2})$. Parameters as in Figure 1. }
	\label{fig:Collapse_revival_mixed_state}
\end{figure}

\begin{figure}
	\centering
\scalebox{0.33}{\includegraphics[angle=0]{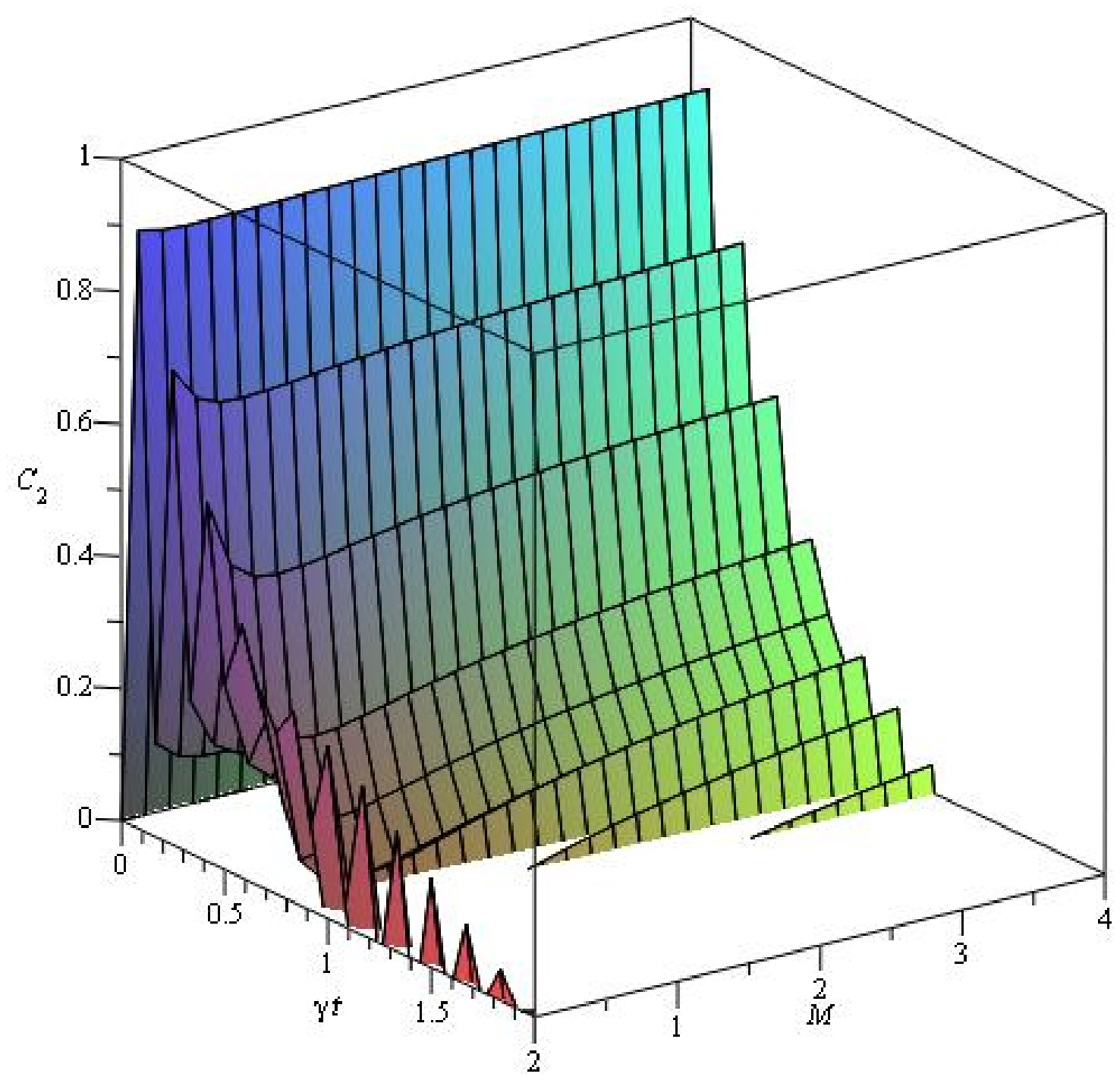}}
\scalebox{0.33}{\includegraphics[angle=0]{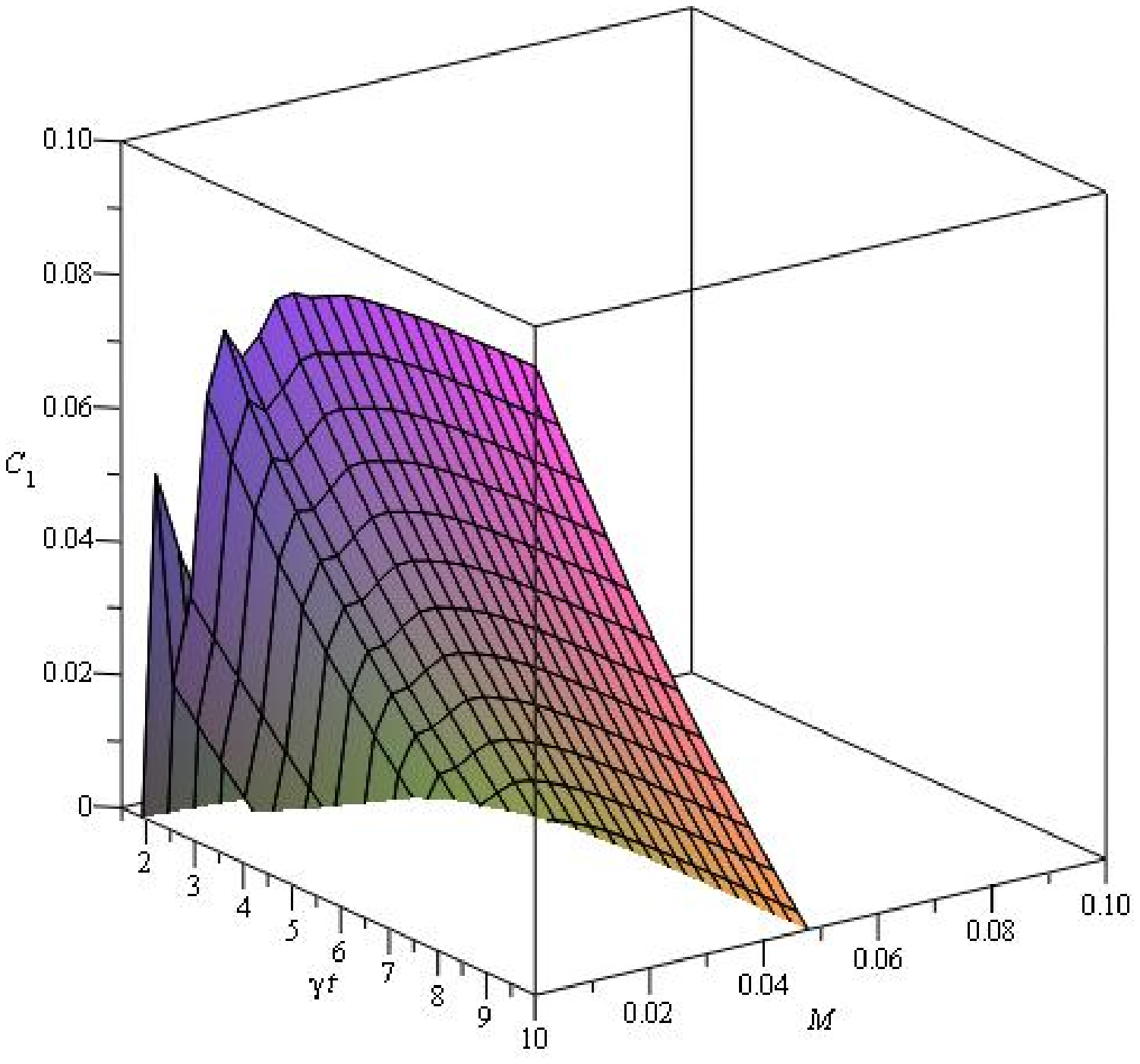}}
	\caption{Concurrence production and collapse (left) and concurrence revival (right) as a function of time and noise parameter M for the product state $\psi_{in}=|eg\rangle$. We observe the stochastic antiresonance behavior in the entanglement oscillations of the left diagram. Parameters as in Figure 1. }
	\label{fig:production_collapse_revival_product_state_eg}
\end{figure}
For initial conditions with matrix representation of the form
\begin{align}
\begin{pmatrix}
0&0&0&0\\0&\bullet&\bullet&0\\ 0&\bullet&\bullet&0\\ 0&0&0&0
\end{pmatrix}
\end{align}
we begin from  the ``central'' subspace (basis $\{|eg\rangle,|ge\rangle\}$) of our density matrix, which for $\gamma\rightarrow0$ becomes a decoherence free subspace \cite{Casati}.
With these initial conditions we studied the  Bell state $\psi_{in}=\frac{|eg\rangle+|ge\rangle}{\sqrt2}$ with matrix representation
\begin{align}
\begin{pmatrix}
0&0&0&0\\0&1/2&1/2&0\\ 0&1/2&1/2&0\\ 0&0&0&0
\end{pmatrix}
\end{align}
the initial mixed state \begin{align}\rho_{in}=1/2|gg\rangle\langle gg|+1/2\Big(\frac{(|{eg}\rangle+|ge\rangle)(\langle{eg}|+\langle ge|)}{2}\Big)\end{align} with matrix representation
\begin{align}
\begin{pmatrix}
0&0&0&0\\0&1/4&1/4&0\\ 0&1/4&1/4&0\\ 0&0&0&1/2
\end{pmatrix}
\end{align}
and the product state $|eg\rangle$ with matrix representation \\

\begin{align}
\begin{pmatrix}
0&0&0&0\\0&1&0&0\\ 0&0&0&0\\ 0&0&0&0
\end{pmatrix}
\end{align}

We present our results for zero temperature in Figures 1 to 6. Figure 1 shows the expected monotonous degradation of entanglement due to external noise for Bell state $\psi_{in}=\frac{|ee\rangle+|gg\rangle}{\sqrt2}$ and product state  $\psi_{in}=|gg\rangle$. In Figure 2 we have the concurrence collapse and revival as a function of time and the noise parameter M for the Bell state $\psi_{in}=\frac{|eg\rangle+|ge\rangle}{\sqrt2}$. We observe the expected collapse and revival of entanglement with the destructive effect of noise. But it is interesting that noise does not affect the concurrence monotonically as can also be seen in Figure 3 for the evolution of linear entropy. In Figure 4 we plot the collapse time and the mixedness as a function of the noise strength, to show this lack of monotonicity more explicitly. We call this negative effect "`stochastic antiresonance"'. In this Figure we also present the dependence of $M_{antires}$  as a function of the parameters $\Delta$ and $\gamma$.  The diagrams show that this dependences is non-zero but very small.  
In Figure 5 we have the concurrence collapse and revival for the mixed state $\rho_{in}=\frac{1}{2}|gg\rangle\langle gg|+\frac{1}{2}(\frac{(|{eg}\rangle+|ge\rangle)(\langle{eg}|+\langle ge|)}{2})$. In Figure 6 we have the concurrence production and collapse and concurrence revival for the product state $|eg\rangle$. It is obvious that noise affects the entanglement production in an unexpected way. The entanglement oscillation amplitudes exhibit stochastic antiresonance behavior. Furthermore we see that for bigger values of M the oscillations last longer.

\section{Concurrence and purity evolution for $T\neq 0$}
For $T>0$ the master equation \eqref{eksisosi} becomes:
\begin{align}
\frac{d\rho_s}{dt}=&\nonumber-i[H,\rho_s]+\gamma( n+1)D[S^-_1]\rho_s+\gamma( n+1)D[S^-_2]\rho_s\\&+\gamma nD[S^+_1]\rho_s+\gamma nD[S^+_2]\rho_s-M[V,[V,\rho_s]],
\end{align}
where $\bar n$ denotes the average excitation quanta of the bath. It depends monotonically on the temperature and it is used to parametrize it. For $T=0$ we have $ n=0$ and for
$T\rightarrow\infty$, $ n\rightarrow\infty$.
The steady state concurrence is found to be equal to:
\begin{align}
&C_{st}=\nonumber\frac{1}{2}\Bigg\{\sqrt {{\frac { \left( {k}^{2}{\gamma}^{2}+ 4kM\gamma+ 4{M}^{2}+{\omega}^{2} \right) {\Delta}^{2}\gamma^2}{ \left( {k}^{3}{\gamma}^{3}+ 4
\,{k}^{2}M{\gamma}^{2}+k{\Omega}^{2}\gamma+ 2{\Delta}^{2}M \right) ^{2}}}}\\&-\sqrt {{\frac { \left( {k}^{2} \left( n+ 1 \right) n{
\gamma}^{3}+ 4M \left(  \left( {n}^{2}+n \right) {\gamma}^{2}+ \frac{1}{2}
{\Delta}^{2} \right) k+ \left({\Omega}^{2}n( {n}+
1)+ \frac{1}{4}{\Delta}^{2} \right) \gamma \right) ^{2}}{{k}^{2} \left( {k}
^{3}{\gamma}^{3}+ 4{k}^{2}M{\gamma}^{2}+k{\Omega}^{2}\gamma+ 2{\Delta}^{2}M
 \right) ^{2}}}}\Bigg\},
\end{align}
where $k=(n+1/2)$ and $\Omega^2=\omega^2+\Delta^2+4M^2$

For given system parameters $C_{steady}$ decays rapidly when noise and thermal excitation parameters increase. Moreover $\bar n$ is a multiplicative factor in the Master equation and degrades the quantum entanglement.

\begin{figure}[htp]
	\centering
\scalebox{0.3}{\includegraphics[angle=0]{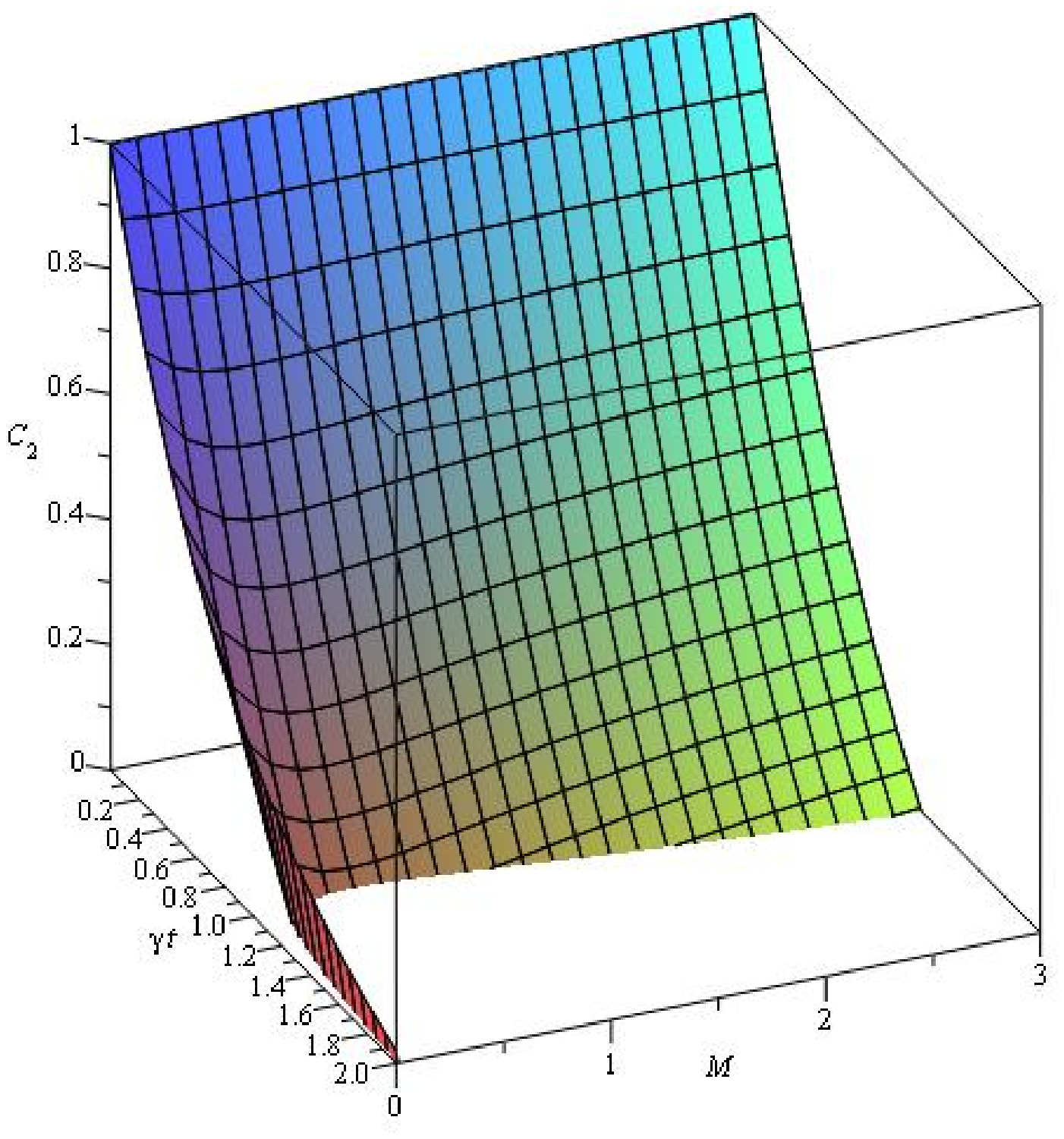}}
\scalebox{0.3}{\includegraphics[angle=0]{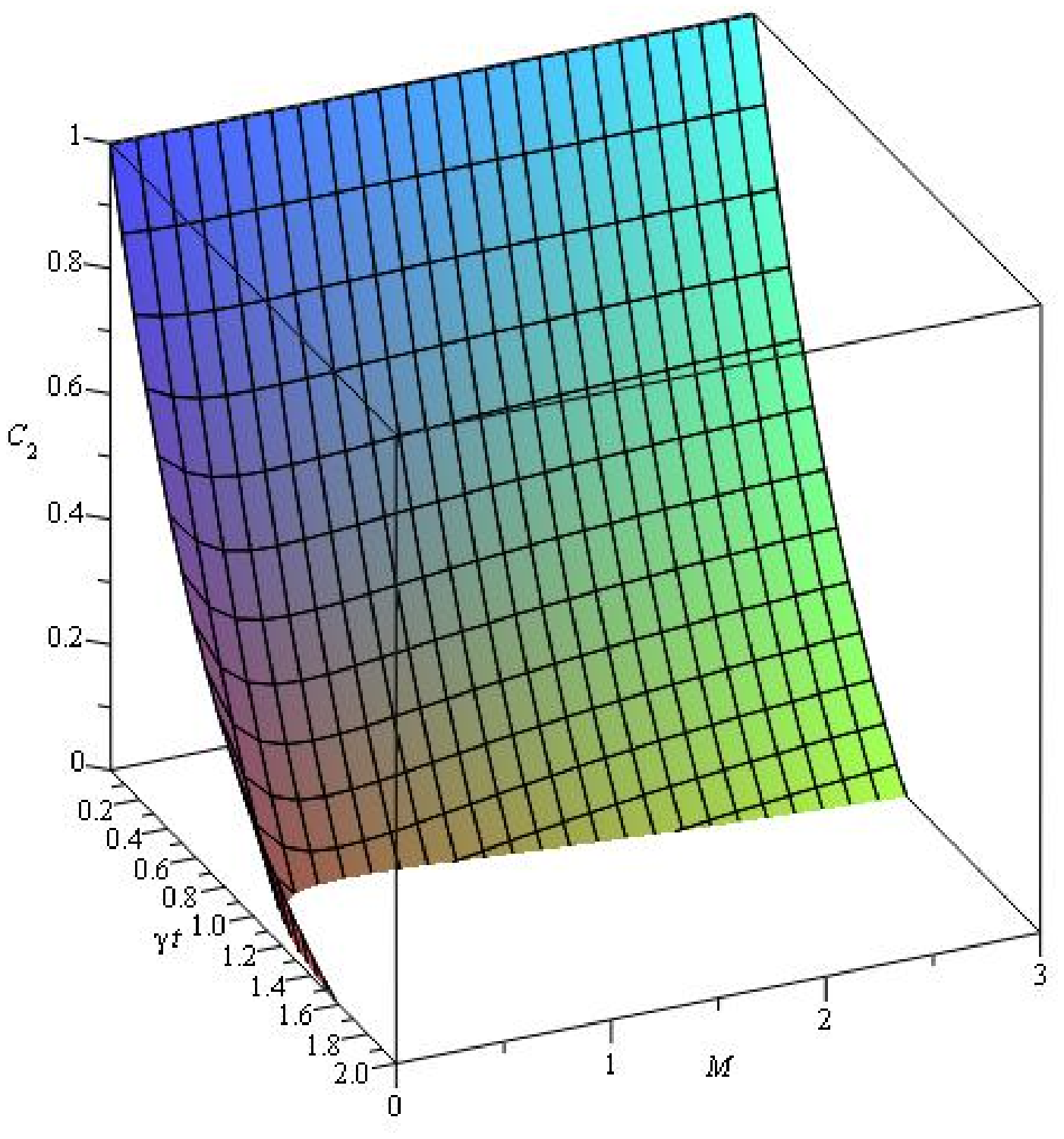}}
\scalebox{0.3}{\includegraphics[angle=0]{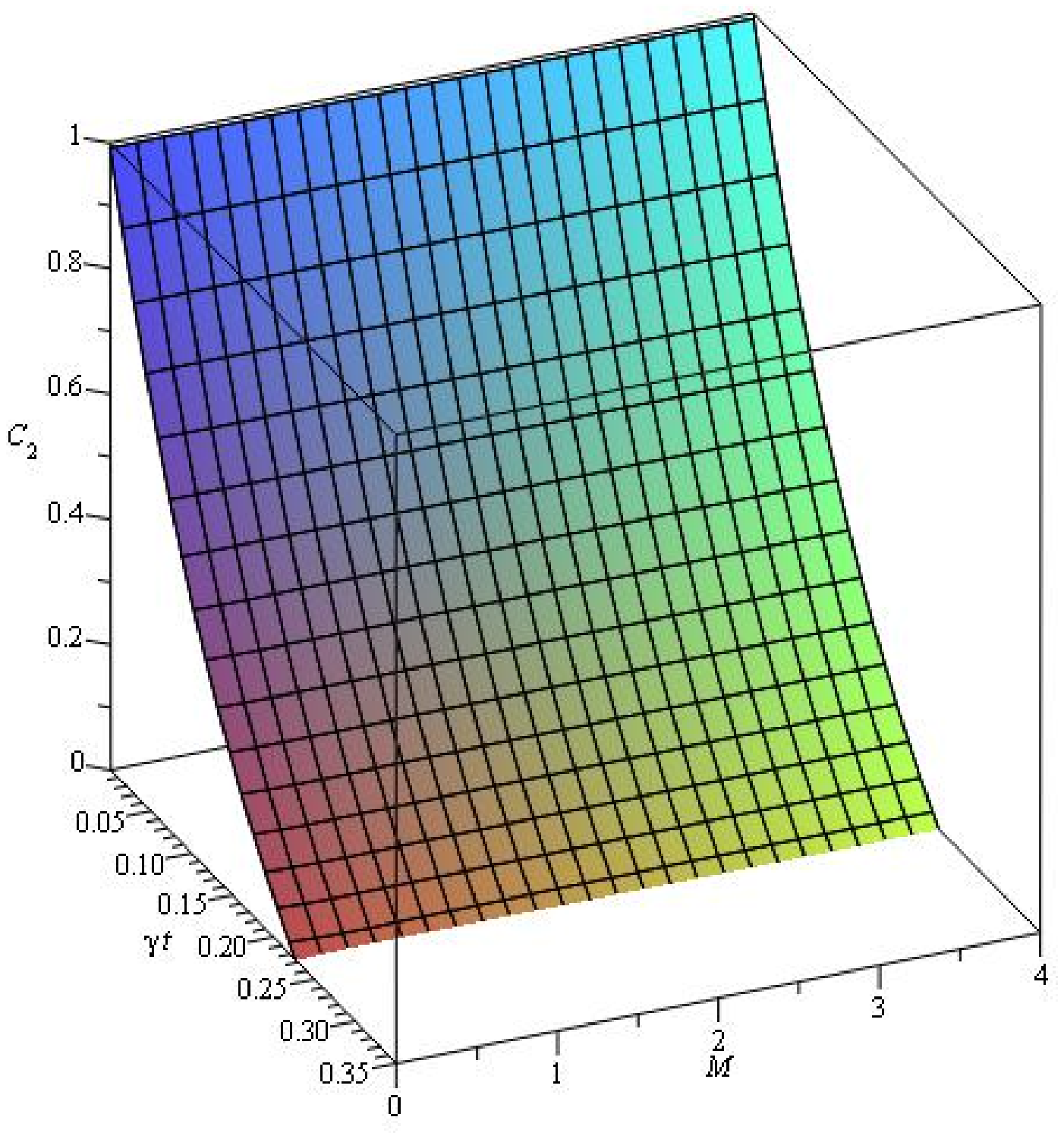}}
	\caption{Concurrence collapse as a function of time and noise parameter M for the Bell state $\psi_{in}=\frac{|eg\rangle+|ge\rangle}{\sqrt2}$ ($J=\Delta=0.1, \gamma=0.01$ and $\omega=1$), for average photon numbers n=0.02, n=0.06 and n=0.1, from left to right. The non monotonicity of the noise influence diminishes progressively.   }
	\label{fig:concurrence collapse}
\end{figure}
In Figure 7 we plot the concurrence collapse as a function of time and noise parameter M for three different values of the average photon number. We observe that the increase of temperature decreases the nonmonotonicity of the noise dependence, something normally expected.
\section{Conclusion}
We studied the entanglement evolution of a two-qubit Heisenberg XY chain in the presence of a noisy magnetic field.  We solved the appropriate Master equation both for $T=0$ and $T\neq0$ and showed that our system presents stochastic antiresonance behavior for certain initial states. These states have matrix representation of the form of equation (20). \footnote{For initial states of this form, the first moments of entanglement evolution (sudden death for initially entangled states and entanglement production for initially product states) are quantified by $C_2$ while the rebirth-steady state entanglement is quantified by $C_1$. }

For $\gamma\rightarrow 0$ the system acquires a decoherence free subspace (DFS), spanned by the basis $\{|eg\rangle,|ge\rangle\}$. Stochastic antiresonance in our system has to do only with the sudden death part of entanglement and not with the rebirth part, something that has advantages and disadvantages.

Noise can play a destructive role in the first moments of evolution. There is a value of noise for which the system has the worst response (the smallest entanglement collapse time) and must be avoided, unlike the stochastic resonance case where we try to reach this value. This can be achieved easily because our noise source is classical and we assume that it is exactly tunable. The above have been confirmed both with the value of the concurrence and that of linear entropy. The steady state concurrence is affected negatively by the external noise. As for the temperature, it affects drastically the system evolution, diminishing the sharpness of antiresonance.

The non monotonic dependence of the disentanglement time, of course makes sense only in the case of a finite time, that is in the case of Entanglement Sudden death (ESD)  \cite{eber1,eber2,eber3,eber4,eber5,eber6,santos,ficek,lastra}. It is interesting to investigate the influence of the geometry of the state manifold \cite{cunha1,cunha2,cunha3} on this interplay of noise, environment and qubit interactions. We are currently looking at these issues.

\pagebreak
\section{Appendix}
\subsection{Exact solution of the master equation}
The differential system that represents our master equation is:
\begin{align*}
&\dot{\rho}_{11}(t)=-2\gamma(n+1)\rho_{11}(t)+(\rho_{14}(t)-\rho_{41}(t))\Delta i+\gamma n(\rho_{22}(t)+\rho_{33}(t))\\&
\dot{\rho}_{14}(t)=(-(2n-1)\gamma-2i\omega-4M)\rho_{14}(t)+(\rho_{11}(t)-\rho_{44}(t))\Delta i\\&
\dot{\rho}_{22}(t)=\gamma(n+1)\rho_{11}(t)-(2n+1)\gamma\rho_{22}(t)+Ji(\rho_{23}(t)-\rho_{32}(t))+\gamma n\rho_{44}(t)\\&
\dot{\rho}_{23}(t)=-(2n+1)\gamma\rho_{23}(t)+Ji(\rho_{22}(t)-\rho_{33}(t))\\&
\dot{\rho}_{32}(t)=-(2n+1)\gamma\rho_{32}(t)-Ji(\rho_{22}(t)-\rho_{33}(t))\\&
\dot{\rho}_{33}(t)=\gamma(n+1)\rho_{11}(t)-(2n+1)\gamma\rho_{33}(t)+Ji(\rho_{32}(t)-\rho_{23}(t))+\gamma n\rho_{44}(t)\\&
\dot{\rho}_{41}(t)=(-(2n+1)\gamma+2i\omega-4M)\rho_{41}(t)-(\rho_{11}(t)-\rho_{44}(t))\Delta i\\&
\dot{\rho}_{44}(t)=\gamma(n+1)(\rho_{22}(t)+\rho_{33}(t))+\Delta i(\rho_{41}(t)-\rho_{14}(t))-2\gamma n\rho_{44}(t)
\end{align*}

We observe that noise enters only in $\dot{\rho}_{14}$  and $\dot{\rho}_{41}$. Indeed, it can be easily shown that for $\gamma\rightarrow 0$, the system acquires a decoherence free subspace (DFS), spanned by the basis $\{|eg\rangle,|ge\rangle\}$. This means that if the initial state of the system resides in the central subspace of the system Hilbert subspace in a symmetric fashion (Bell states $\frac{|eg\rangle\pm|ge\rangle}{\sqrt2}$), then it remains the same. This central subspace of the general density matrix plays a crucial role for the unexpected resonance like behavior of certain initial states.

The introduction of noise in the master equation makes the general solution quite complex. In the steady state limit the solution of our system is:

\begin{align*}
&{\rho}_{11}(t)=\frac{1}{16}{\frac {\left( 4\gamma n+\gamma+8Mk \right) {\Delta}^{2}+4{n}^{2}\gamma
 \left( {k}^{2}{\gamma}^{2}+4Mk\gamma+{\Omega}^{2} \right) }{ \left( {k}^{3}{\gamma
}^{3}+4M{k}^{2}{\gamma}^{2}+{\Omega}^{2}k\gamma+2M{\Delta}^{2} \right) k}}
\\&
{\rho}_{14}(t)=-{\frac { \left( 2i\gamma n+4iM+i\gamma+2\omega \right) \gamma\Delta}{8{k}^{3}
{\gamma}^{3}+32{k}^{2}M{\gamma}^{2}+8{\Omega}^{2}k\gamma+16M{\Delta}^{2}}}\\&
{\rho}_{22}(t)={\frac { \left( \gamma+8kM \right) {\Delta}^{2}+ \left( {n}^{2}+n
 \right)  \left( 4{k}^{2}{\gamma}^{3}+4\gamma \left( {\Omega}^{2}+4kM\gamma
 \right)  \right) }{16{k}^{2}\gamma \left( {k}^{2}{\gamma}^{2}+4kM\gamma+{\Omega}
^{2} \right) +32k{\Delta}^{2}M}}
\\&
{\rho}_{23}(t)=0\\&
{\rho}_{32}(t)=0\\&
{\rho}_{33}(t)={\rho}_{22}(t)\\&
{\rho}_{41}(t)={\rho}_{14}^*(t)\\&
{\rho}_{44}=1-{\rho}_{11}-{\rho}_{22}-{\rho}_{33}
\end{align*}
where $k=(n+1/2)$ and $\Omega^2=\omega^2+\Delta^2+4M^2$. In this case $C=C_{1}$, because ${\rho}_{32}(t)=0$ and $C$ cannot take negative values.

\end{document}